\documentclass[a4paper,12pt]{article}

\usepackage[a4paper,margin=2cm,marginparwidth=1cm]{geometry}
\usepackage{amsmath,amssymb}
\numberwithin{equation}{section}

\usepackage{graphicx}% Include figure files

\usepackage{hyperref}

\graphicspath{ {images/} }
\usepackage{enumerate}
\usepackage{cleveref}%%cross refer multifigures
\usepackage{graphicx}
\usepackage{subcaption}
\usepackage{caption}
\usepackage{inputenc}
\newcommand{\tcomment}[1]{\textbf{(T)} #1 \textbf{(end of T)}} %{}

\newcommand{\figref}[1]{Fig.\ \ref{#1}}
\newcommand{\Figref}[1]{Fig.\ \ref{#1}}

\providecommand{\citeauthor}[1]{\texttt{#1} \cite{#1}}

\newcommand{\kin}{k^{\mathrm{(in)}}}

\newcommand{\matrx}[1]{{\mathbf{\textsf{#1}}}}

\newcommand{\adj}{\matrx{A}}
\newcommand{\Tmat}{\matrx{T}}

\newcommand{\vertex}{\mathcal{V}}

\begin{document}

%%%%%%%%%%%%%%%%%%%%%%%%%%%%%%%%%%%%%%%%%%%%%%%%%%%%%%%%%%%%%%%
%
% TITLE PAGE
%
% This is a hand built version.
% Each journal may have its own format.
%
% \begin{flushright}
% \textbf{Confidential DRAFT}. \\ Not to be redistributed without permission of the author.\\
% \texttt{Imperial/TP/19/TSE/?} \\
% %\eprint{arXiv:1903.03667} \\
% %14th June 2019 \\
% %\tsecompldate \\
% (\texttt{\jobname.tex}  LaTeX-ed on \today ) \\
% %{next preprint number}\\
% \end{flushright}
% \vspace*{0.5cm}

\begin{center}
{\Large\textbf{Emergence of community structures through biased random walks rewiring}}\\[6pt]
 {\large
 Qing Yao,
 Bingsheng Chen,
 \href{http://www.imperial.ac.uk/people/t.evans}{Tim S.\ Evans},
 Kim Christensen
 }
\\[6pt]
{Centre for Complexity Science and Blackett Laboratory},
Imperial College London, South Kensingtom Campus, London SW7 2AZ, United Kingdom
\\[6pt]
\today
 %\\  Key Words: Graph Theory, History of Mathematics
\end{center}

%
% END OF TITLE PAGE
%
%%%%%%%%%%%%%%%%%%%%%%%%%%%%%%%%%%%%%%%%%%%%%%%%%%%%%%%%%%%%%%%

%\title{Emergence of scaling modular structures through random walks rewiring}% Force line breaks with \\

%\author{Qing Yao}
%\author{Bingsheng Chen}
%\author{Tim Evans}
%\author{Kim Christensen}



\begin{abstract}
Community structures have been identified in various complex real-world networks, for example, communication, information, internet and shareholder networks. The scaling of community size distribution indicates the heterogeneity in topological structures of the network. The current network generating or growing models can reproduce some properties, including degree distributions, large clustering coefficients and communities. However, the scaling behaviour of the community size lacks investigation, especially from the perspectives of local interactions. Based on the assumption that heterogeneous nodes behave differently and result in different topological positions of the networks, we propose a model of designed random walks in directed networks to explain the features in the observed networks. The model highlights that two different dynamics can mimic the local interactions, and a hidden layer is essential when reproducing the characteristics of real complex networks. The key features the model can explain include community size distribution, degree distribution, percolation properties, distribution of average path length and dependence of the above properties on the labels of nodes in the data.
\end{abstract}

\section{Introduction}
Complex networks are informative abstractions of complex systems varying from economic systems to biological systems. A community in a network is defined as a group of nodes connected more densely than the nodes outside the community. Newman places firm grounds on communities which are useful structures of complex networks~\cite{newman2003structure}. Then, many real-world phenomena display not only a scaling behaviour of the degree distributions but also a scaling behaviour of the community sizes distributions~\cite{lancichinetti2010characterizing}, see~\Figref{fig:com_dis}. Similar behaviours are observed for shareholder networks for Netherlands and Turkey~\cite{yao2019network}. The universal behaviours of the complex networks may emerge due to some common mechanism. Different models have been proposed to study the generation of networks. The work of Bogu{\~n}{\'a} \textit{et al.} proposes a model based on social distance~\cite{boguna2004models}. The Shrinking Diameters Model proposed by Leskovec \textit{et al.} is based on `forest fire' and community guided flavours~\cite{leskovec2007graph}. The equitable random graphs introduced by Newman and Martin, give properties discovered in real networks, including communities structures~\cite{newman2014equitable}.

\begin{tabular}{ccc}
     \includegraphics[width=0.29\textwidth]{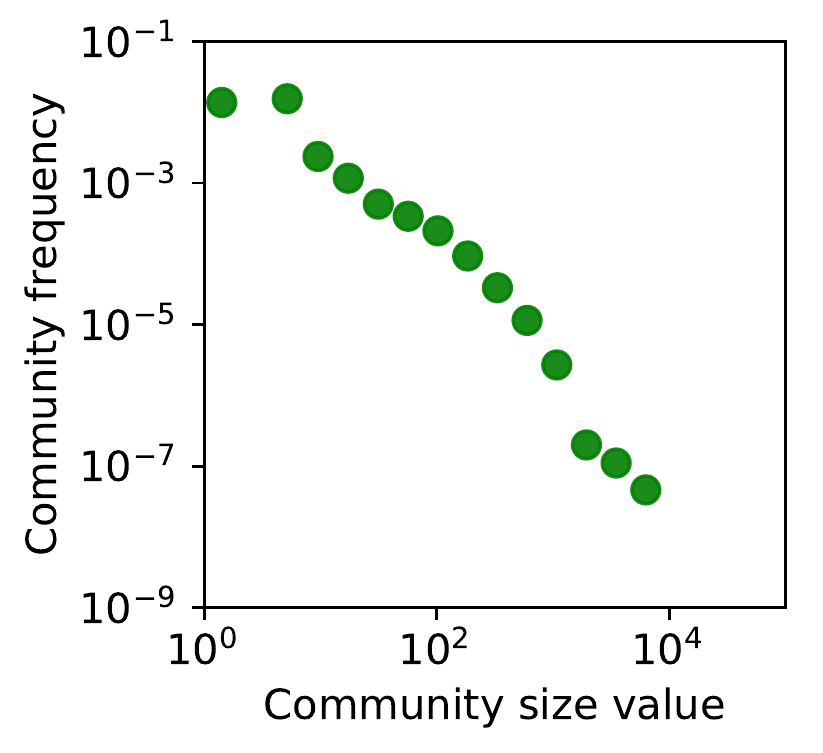} &
     \includegraphics[width=0.3\textwidth]{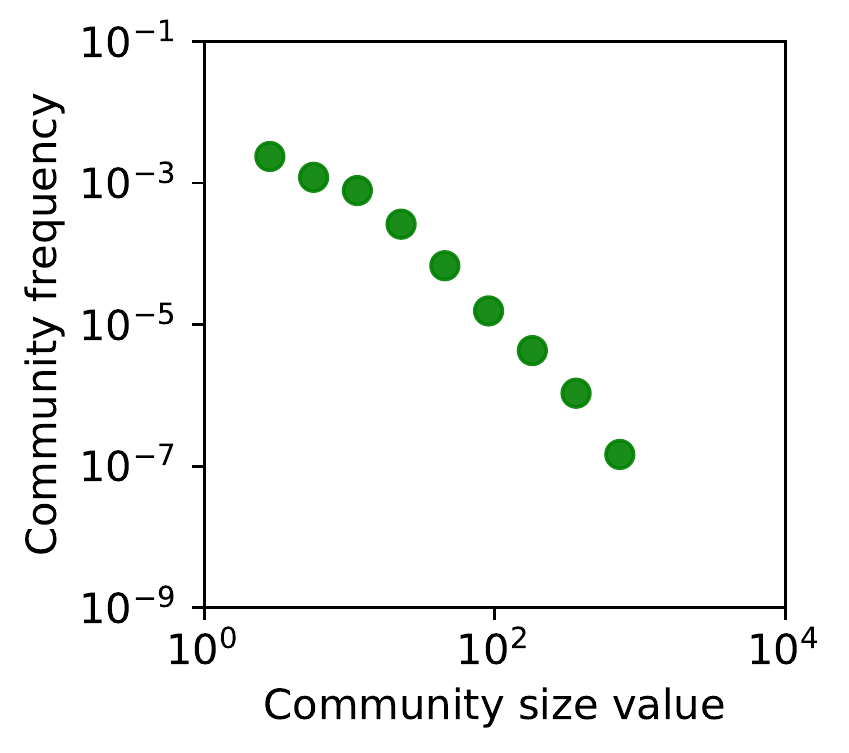} &
     \includegraphics[width=0.3\textwidth]{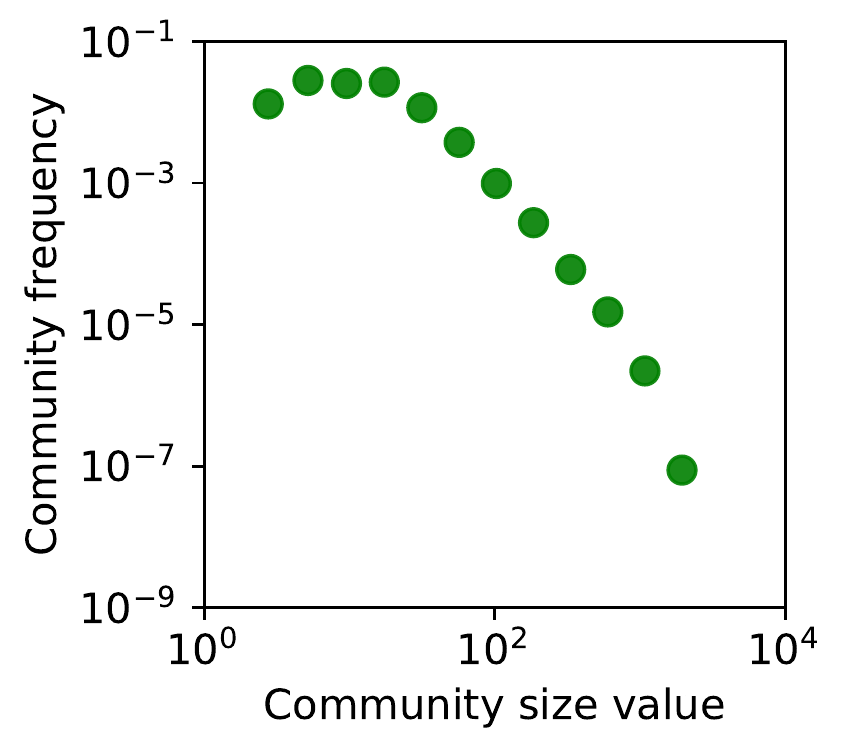}\\
     (a) Communication-Email &
     (b)  Internet-AS-caida &
     (c)  Information-Amazon
\end{tabular}

\begin{figure}[h!]
    \centering
    % Fine to use this in thesis but avoid this in a journal
    %\includegraphics[width = \textwidth]{communities_size}
    %QY to provide own figure generated from data for one or two examples e.g. arXiv and amazon example
    \caption{Distribution of community sizes detected by Infomap for three complex real-world networks. All distributions are broad, and similar for systems in the same category. Data points are
averages within logarithmic bins of the community size. }
    \label{fig:com_dis}
\end{figure}
% The other network characteristics is in the supplementary materials.

The above models can reproduce some of the properties of a real-world network, for instance, fat-tailed degree distribution, high clustering coefficients, and scaling in community size distributions. However, local searching or interaction have not been discussed. To account for local interactions, we propose a different model, which is based on the random walks rewiring of networks consisting of different types of nodes. We show that the model reproduces the main statistical characteristics of real-world systems: the emergence of scaling of communities and percolation properties, the distribution of average shortest path and the dependence of above properties on the types of nodes in the data, especially the results in the work of Shareholder networks~\cite{yao2019network}.

This paper first illustrates how to perform a random walk rewiring process on a directed Erd\H{o}s R\'enyi graph and project onto an undirected graph. Then we introduce a non-topological attribute `type' of the nodes into the rewiring process to account for the nodes' heterogeneous behaviours. Finally, we discuss the results of the model. This model corroborates our intuition that random walk rewiring can mimic the real world; that type is essential when modelling network dynamics and should consider the directions that might imply hidden information.

\section{The model}\label{sec:level1}

\subsection{Random walk rewiring on directed networks}\label{sec:implementation}

Random walk rewiring model uses the random walker to search for the target for the rewiring. The random walk searching can manifest the fundamental concept of cyclic closure~\cite{kossinets2006empirical}. Cyclic closure bias is defined as the empirical probability that two previously unconnected nodes which are a distance apart in the network will initiate a new tie. Thus cyclic closure naturally generalises the notion of triadic closure, i.e., the formation of cycles of length three. From the sociological perspective, Georg Simmel proposed the triad closure concept, and then Mark Granovetter made this concept popular in the paper `The strength of weak tie' ~\cite{granovetter1977strength}. From the aspect of information searching, it is unlikely for people or organisations to have full access to the information of the system in which they are embedded. It is more realistic that a person may follow the choices of the people who are around that person. `Following the strategies of other people' has been successfully applied to recommendation systems, proving that an individual's social information will improve the recommendation system from the collaborative filtering techniques, ~\cite{o2005trust,konstas2009social,ma2011recommender}.

A classic random walk on a graph is often used to represent the way people search a network to find new connections. Node A is connected to node B, which in return is connected to node C, and this suggests that it is natural for node A to connect to node C. However, in many cases, this does not capture the variety of interactions and the different possible choices made in practice. For example, node A and B may be friends because they grew up together while nodes B and C are friends because they work together, yet people rarely introduce their intimate friends to their work colleagues.

In the example of shareholder networks~\cite{yao2019network}, we first created a directed graph with links representing shareholders investing in companies. Then we analysed the shareholder networks where links represent the common companies that shareholders hold in common. To understand the emergence of characteristics of the shareholder networks, a random walk on a larger directed network can represent a wider variety of interactions, particularly when we use different biases in the random walks we use to capture different individual choices.

\begin{figure}[!h]
\centering
\includegraphics[scale =0.9]{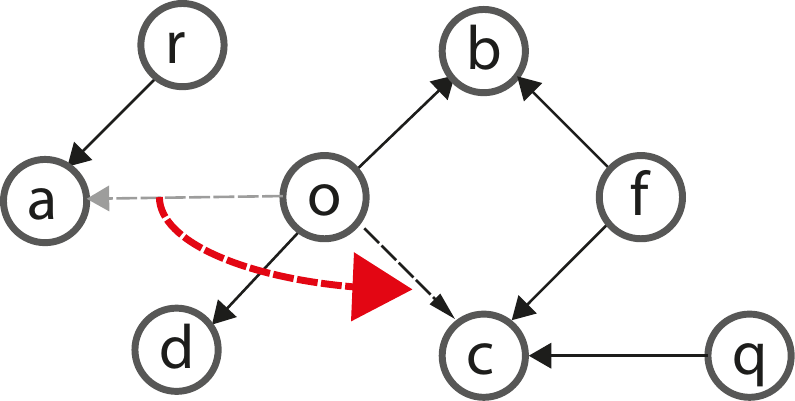}
\caption{The illustration of the \textbf{RRWD} model, rewiring based on the random walk on a directed network. The directed edge $d_{oa}$ is rewired to $d_{oc}$ based on the random walk starting from $o$.}
\label{fig:illu_rw_dr}
\end{figure}

Based on the above argument for a wider variety of interactions, the model we propose starts from a directed graph. In the directed graph, we consider the direction of arrows as the direction of information flow and the nodes as sources and receivers at the same time. The action a node will follow is shown in~\autoref{fig:illu_rw_dr}. If a node $o$ as a source wants to find a new receiver it will use a simple random walker following the direction of information flow to find one of its current receivers, say, $b$. Then it makes a second step back to find one of the sources of $b$, say node $f$. The backwards step allows our original node $o$ to find ancestor node $f$ with a similar relationship to common node $b$. For example, $o$ and $f$ are both shareholders in company $b$. Finally, we find a different receiver node of $f$, node $c$. The node $c$ is a possible new receiver for $o$ because we assume if two nodes have a similar relationship to one common node, they are likely to have relationships to more nodes. That means $o$ may want to follow $f$'s relationship to $c$.

For the original node, $o$, which can now make a new connection, the detailed procedures are in the following:
\begin{enumerate}
\item Initialisation with an Erd\H{o}s R\'enyi directed graph $\mathcal{D}(N, p)$, where $N$ is the number of nodes, $p$ is the probability for creating one edge between a pair of nodes (direction assigned randomly) and $m$ is the number of edges in that graph; $\mathcal{D}$ has a set of nodes $V = \{a, b, c, d...\}$, with $|V| =  N$ members.
\item \label{picknode} Pick randomly a node $o \in V$, whose degree, $k_{o} \geq 2$, that node $o$ has more than one neighbours. Next randomly follow an edge from $o$ to one of its neighbours (in this case $a$). This edge will be rewired.

\item \label{startwalk} A random walker starts from $o$ and walk to a neighbour vertex $b$. Following the direction of the edge, it can not walk to $a$.

\item Let the walker walk to an ancestor of $b$, node $f$, so going against the direction of the edge.

\item Let the walker walk to a neighbour of $f$, node $c$, now following the edge direction.

\item Check if the edge $(o, c)$ exists in the graph. If not, delete the edge $(o, a)$ and create a new directed edge $(o, c)$. If the edge $(o, c)$ already exists in the graph, make node $c$ as the starting node and follow step 3, until a new edge is found or exceeds the maximum trial $(100)$.

\item Go back to step 2.
\end{enumerate}

We call this rewiring based on random walk on directed graph, \textbf{RRWD} model in the following context.

\tcomment{We need a clearer definition of a network here.  I think it is bipartite not just directed. Q: The literature shows that any directed graph has a corresponding bipartite graph. i think the bipartite corresponding graph just help the explanation. However, the labelling of the bipartite is not unique if corresponding adjacency matrix $A(B)$ constructed from the $A(D)$ is disconnected.}

\begin{equation}
    \adj(B) = \begin{bmatrix}
        0 & \adj(D) \\
         \adj^{t}(D) & 0
        \end{bmatrix}
\end{equation}

The random walks on a directed unweighted graph, the transition probability $\Tmat_{ij}$ of a walker starting from node $i$ and landing at node $j$ can be written as:
\begin{equation}
    \Tmat_{ij} = \frac{\adj_{ij}}{k^{out}_{i}} \,
\end{equation}
where $A_{ij}$ is the adjacent matrix which is changing with the time and $k^{out}_{i}$ is the number of out-going edges of node $i$~\cite{masuda2017random}.

If we define $P_{i}(t)$ as the probability node $i$ is visited at discrete time $t$, then the probability node $j$ is visited at time $t+1$ is
\begin{equation}
  P_{j}(t+1) = \sum^{|\vertex|}_{i = 1}P_{i}(t)\Tmat_{ij},
\end{equation}
where the sum runs over all the nodes. If $\mathbf{T}$ denotes the transition matrix with the entries $\Tmat_{ij}, i, j = 1,2,...,|\vertex|$.

\subsection{Projection of the directed network} \label{subsec:pro}
\begin{figure}
\centering
\includegraphics[scale=0.9]{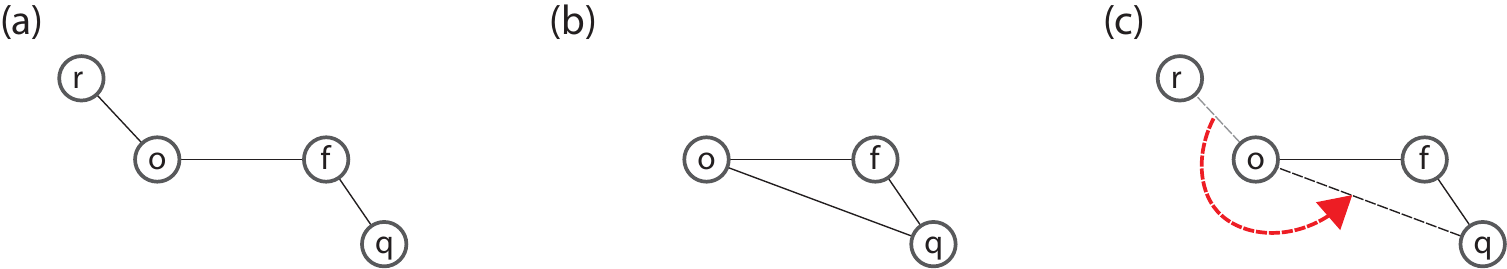}
\caption{The illustration of the undirected networks projected from the rewired directed networks. After rewiring, the edge undirected edge $e_{r0}$ is rewired to $e_{oq}$. A triangle between nodes $c$, $f$ and $q$ is created.}
\label{fig:illu_rw_pj}
\end{figure}

Then, an undirected network is constructed based on the common heads of arrows in the directed network: if two nodes have one common target, as in Figure~\ref{fig:illu_rw_dr}, nodes $p, q$ and $o$ have one common target $c$, in projected network Figure~\ref{fig:illu_rw_pj}, an edge is created between $f$ and $q$, $f$ and $o$ and $o$ and $q$.

When projecting the directed graph, if a node has a number of indegrees $k$, it will create a clique of $k$ fully connected nodes. To lease the effect of these types of cliques on community detection, we choose to use a weighted graph when projecting the graphs. The directed graph is $\mathcal{D}(V, A)$, where $V$ is a set of nodes with outgoing edges and $A$ is a set of nodes with incoming edges. $|V \cup A| = N$. In the projected graph $\mathcal{P}(V,E)$, an edge $e_{ij} \in E$ exists when in $\mathcal{D}$, nodes $i, j \in V$ have a common target $c \in A$. The weight $w_{ij}$ of an edge $e_{ij}$ is the sum of the weights of common targets, for one common target $c$ and the $w_{c}$ is the inverse of the number of edges incoming of $c$. In terms of the adjacent matrix,
\begin{equation}
  P_{ij}
  =
  \begin{cases}
  \sum_{c}\frac{2}{\kin_c(\kin_c -1)} & \mbox{if }  \kin_c>0\\
  0 & \mbox{otherwise }
  \end{cases}
  \, ,
  \label{Pmatdef}
\end{equation}
where $\kin_c$ is the in-degree of the common target node.

\subsection{Two types of vertices}\label{subsec:label}
When considering random walk rewiring, we interpret the description `following the random walker on the graph' as behaviours of following other people around you. Based on this interpretation, we can consider the constraints depending on in-edges as individuals who have niche tastes and who are not willing to get attached too `popular' targets.
In the real world, different types of individuals or organisations have initiatives to make different choices. To mimic this heterogeneous behaviour, we assign `types' to the nodes when we initialise the graphs and let different types of vertices have preferences of in-edges of the target nodes when making a rewiring. The types of nodes do not change during the process, and the node where a random walker starts is also the source of the edge to be rewired.

\begin{enumerate}
\item Type~$1$, prefers to attach to a node with more edges than $\theta_{e}$. In the real world, this type of nodes would be the population following the majority taste, or the investors who would like to hold the assets sharing risks with others.
\item Type~$2$, prefers to attach to a node with less than or equal to $\theta_{e}$ edges. In the context of the real world, this can be the type of person who has a niche taste, or the investor who would like to control the invested company.
\end{enumerate}

For simplicity, in this rewiring based on biased random walk on directed graph with labelled nodes(\textbf{RBRWD}) model, we take two types of vertices into account and the threshold of edges $\theta_{e}= 2$, including the edge to be attached. However, more types can be defined, and the interaction among them can be inferred from the real data.

\begin{figure}
\centering
\includegraphics[scale=0.9]{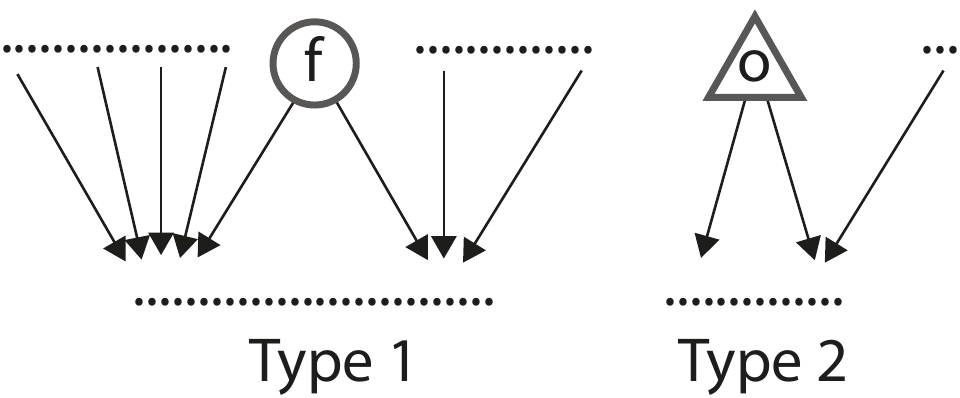}
\caption{Illustration for two types of vertex, type 1 prefer to attach to targets with lots of other predecessors; type 2 prefer to attach to targets with no more than two predecessors, that is $\theta_{e} = 2$.}
\end{figure}

Networks of typed nodes (the labelled networks) and biased random walks are discussed in the work of Lambiotte \textit{et al.}~\cite{lambiotte2011flow}. We can extend our expression for directed networks to $G(V,A,X)$, where $X = x_1, x_2,...x_{N}$ is the attribute variable of the network. $x_i$ is drawn from the set of two types of nodes {0,1}, that is $x_i =0$ if node is Type $1$ and $x_i = 1$ if node is Type
$2$. Then we can define the coefficient of a biased random walker as $\omega_{ij}$ for the walker from node $i$ to node $j$
\begin{equation}\label{eq:w_theta}
    \omega_{ij} = x_{i} \Theta( \theta_{e}-k_{j}) + (1-x_i)(1- \Theta( \theta_{e}-k_{j})),
\end{equation}
where $\Theta$ is a step function of variable $(\theta_{e}-k_{j})$: $=1$ if $k_j \leqslant \theta_{e}$ and $=0$ if $k_j > \theta_{e}$.

Then the transition probability from node $i$ to node $j$, $\Tmat_{ij}$ would be modified as:
\begin{equation}
    \Tmat^{\omega}_{ij} = \frac{\omega_{ij}\adj_{ij}}{\sum_{k}\omega_{ik}\adj_{ik}},
\end{equation}
where $\adj$ refers to the adjacency matrix. The equilibrium probability landing on node $j$ has not been solved.

\begin{figure}
    \centering
    \includegraphics[scale=0.8]{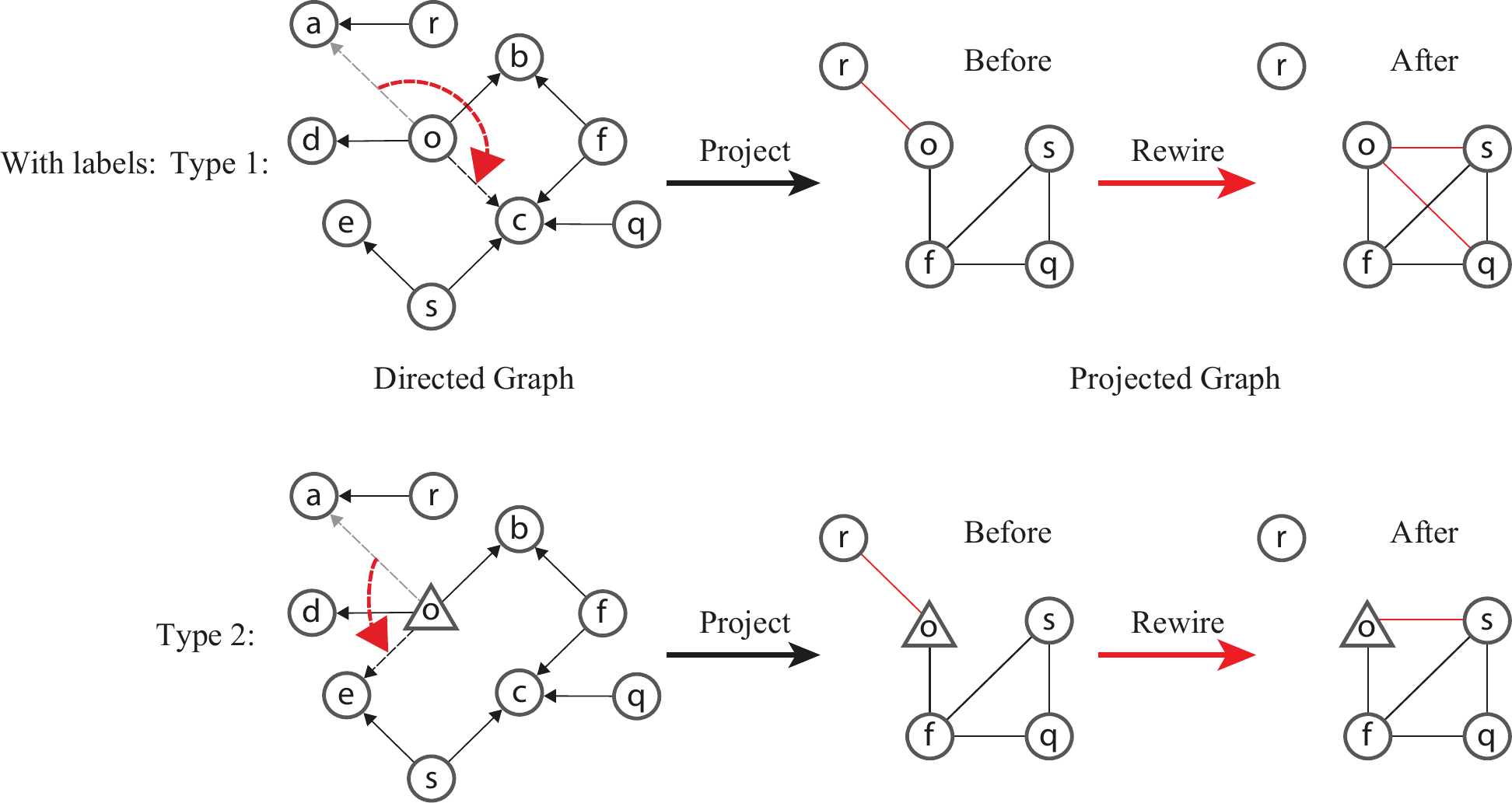}
    \caption{Illustration of \textbf{RBRWD} model, rewiring based on the random walk on a directed network with labelled nodes. If node $o$ is a type 1 node, the directed edge $d_{oa}$ is rewired to $d_{oc}$ based on the random walk starting from $o$. If node $o$ is a type $2$ node, and therefore prefers to attach to nodes with fewer predecessors $(\theta_{e} = 2)$, then when random walker reaches $c$, who has 3 predecessors $(>\theta_{e})$, continues walking to $s$ and $e$. The latter  node $e$ has only one predecessor $(<\theta_{e})$ and the rewiring is performed.}
\end{figure}

%\section{\label{sec:implementation}Implementation}
\subsection{Steady State and Randomness}
This model preserves the number of edges in the directed network. By measuring average clustering coefficients and the number of edges in the projected networks, we can observe if the system has reached a steady-state. %The x-axis now is plotted as the generation of the directed networks.
A ``generation time'' is defined as the number of nodes in the networks. For each system size $|\vertex|$, the generation is different. In the steady-state, the node will not always be in the same community, while communities statistics stay stable.

%\subsection{Randomness}
Besides following others and the niche tastes, there are some other factors, like the information shared via the social media platform. To make the process capture the other rewiring mechanism, we add randomness to the model. When a node is picked up, and with a probability of $P_{rw}$, a random walk will be performed, with a probability of $P_{r} = 1-P_{rw}$, a rewire will be performed instead to a random target.

% \section{more results}
% \begin{figure}
% \centering
% \includegraphics[scale=0.1]{./images/g000p_label_200_098_V3}
% \caption{The initial projected network of $200$ nodes and $993$ edges. The clustering coefficient is $0.44$. There are equal numbers of two different types, Type $1$ (circle) and Type $2$ (triangle).}
% \end{figure}

% \begin{figure}
% \centering
% \includegraphics[scale=0.09]{./images/g0p_label_200_098_V3}
% \caption{The projected network in the steady state with $200$ nodes and $1315$ edges, using $P_{rw} = 0.1$. The clustering coefficient is $0.59$. Type $1$ (circle) and Type $2$ (triangle). Colours represented different communities detected using Infomap. }\label{fig:results0}
% \end{figure}

% *****************************************
\section{Results}\label{sec:result}

\begin{figure}[!h]
    \begin{tabular}{cc}
     \includegraphics[scale=0.09]{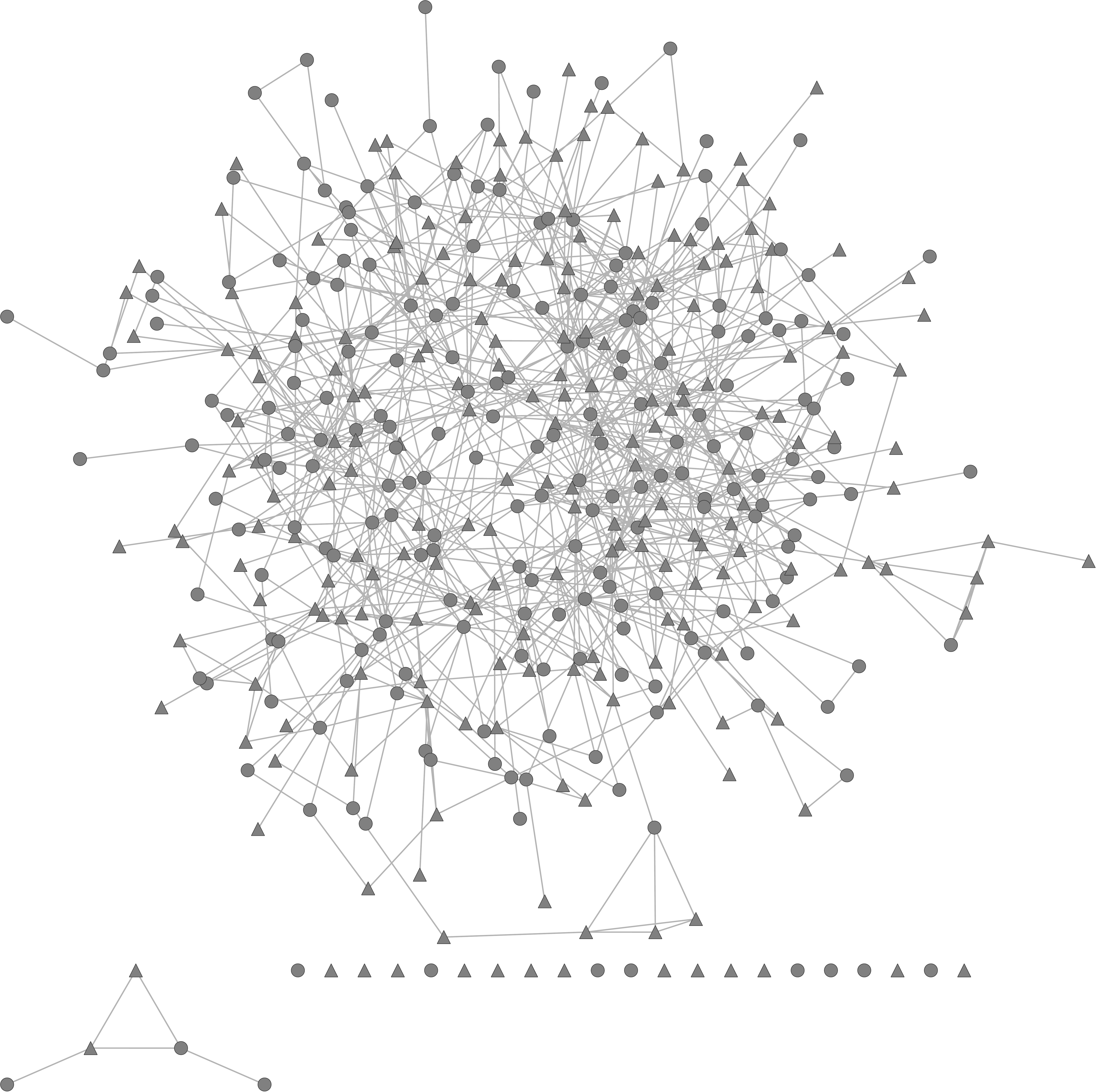}    &
     \includegraphics[scale=0.07]{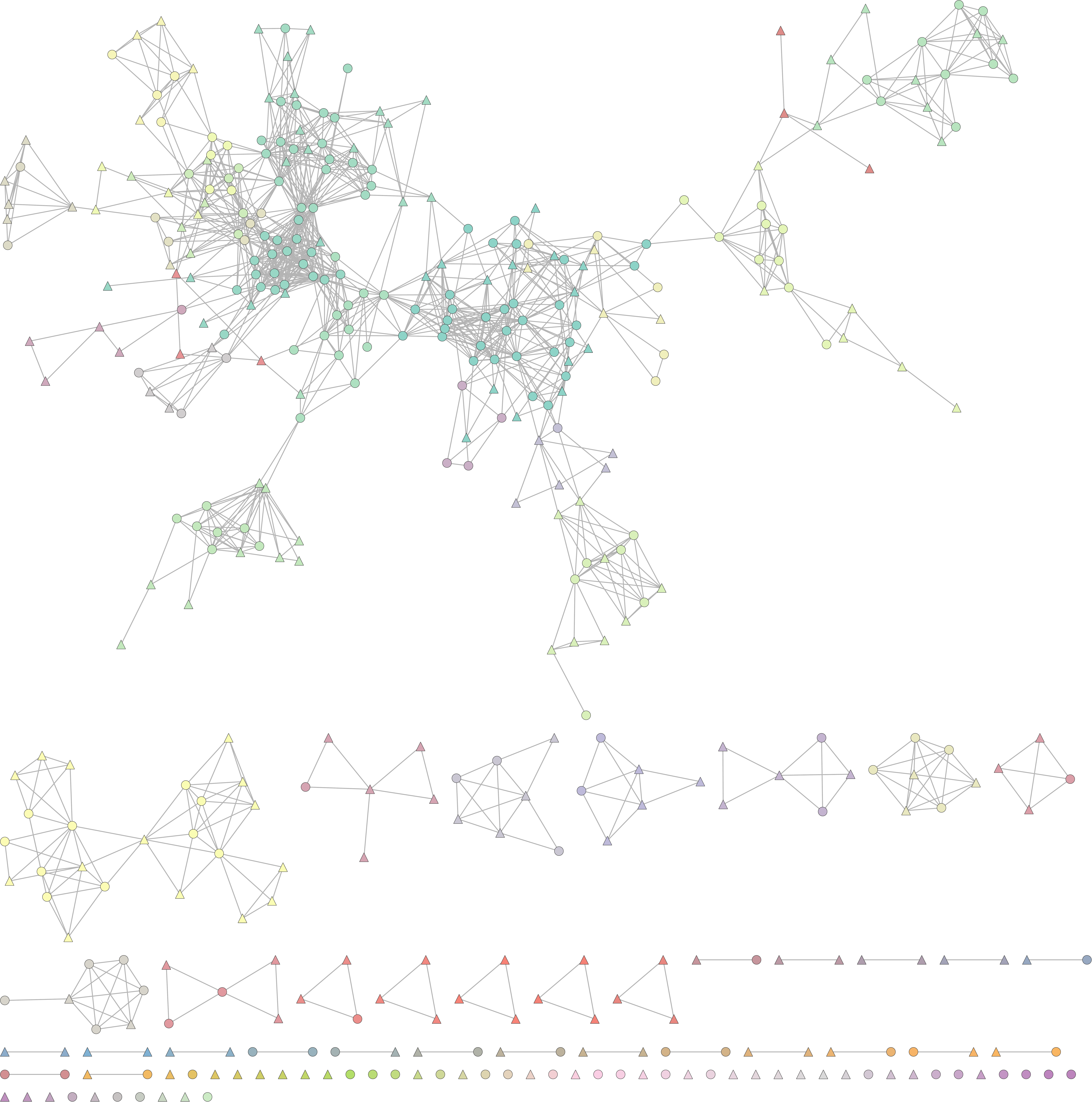} \\
     Projected network of initial graphs. & Projected network after 4 `generation time evolving'.
    \end{tabular}
    \caption{The projected network of initial random graph has $200$ nodes and $993$ edges. After 4 'generation time' evolving with $P_{r} = 0.1$, The projected network has $200$ nodes and $1315$ edges. The clustering coefficient is $0.44$. There are equal numbers of two different types, Type $1$ (circle) and Type $2$ (triangle).}
    \label{fig:results0}
\end{figure}

The result in \Figref{fig:results0}, shows that a giant component emerges from the randomised graph in \textbf{RBRWD} model. Small components with two types of nodes and isolated nodes are scattered around.

In \textbf{RBRWD} model, once a random directed graph is created, we fix the number of edges.
The parameters are the number of nodes $|\vertex|$ and the edge creation probability $p_{er}$ in the directed graph(equally the expected number of edges, since in ER model, $\left<m\right> = \binom n2 p_{er}$). When $|\vertex|$ is large, $p_{er} \approx \frac{4}{|\vertex|}$ and $P_r$ in the range of $(0.1, 0.01)$, The fat-tail distribution of community size is expected. The fraction of different types will not affect the form of communities, i.e., when applying to rewire heterogeneously, the scaling of community size will always emerge.

In what follows, we compare the measurements of the projected graphs, $\mathcal{P}$ with what we have observed in real data.

% -----------------------------------------------------
\subsection{The community distribution}

\begin{figure}[!h]
\begin{tabular}{r||c|c|c}
$|\vertex|$ & Initial & Without labels & With labels\\ \hline\hline
\raisebox{0.15\textwidth}{3000}
& \includegraphics[width=0.25\textwidth]{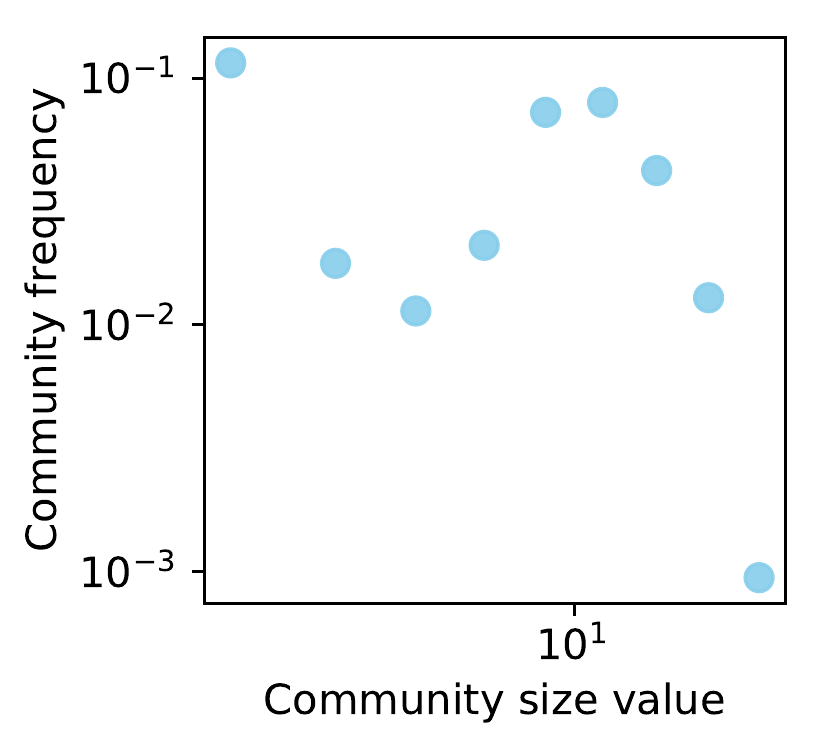} & \includegraphics[width=0.25\textwidth]{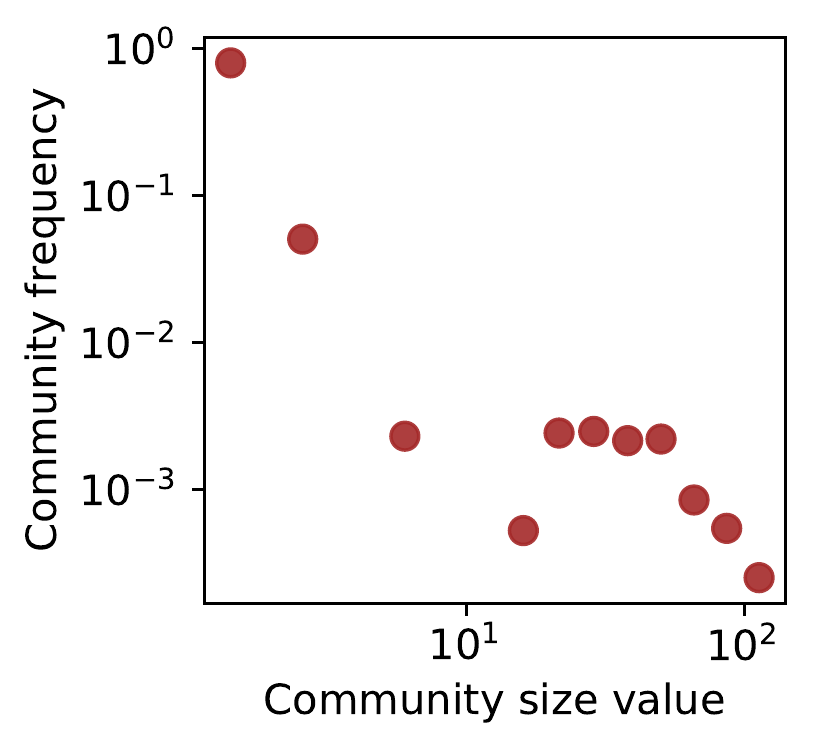} &
\includegraphics[width=0.25\textwidth]{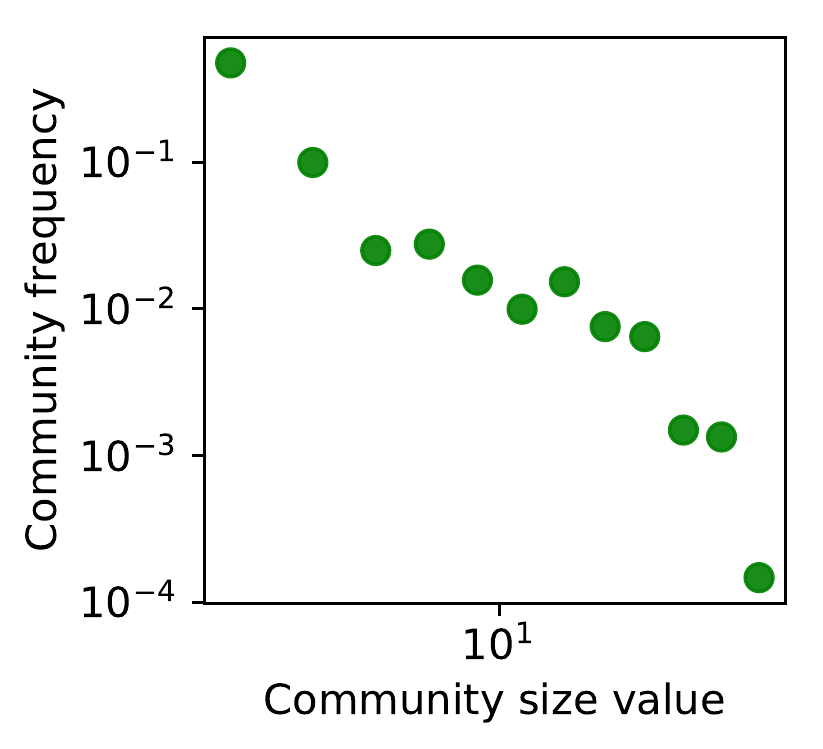}\\
$a$ bin size value & $1.25$ &$2$ or $1.35$ &$2$ or $1.35$ \\
$p$ value & $0.08$ & $0.31$ or $0.15$ & $0.81$ or $0.31$ \\ \hline
\raisebox{0.15\textwidth}{40000}
& \includegraphics[width=0.25\textwidth]{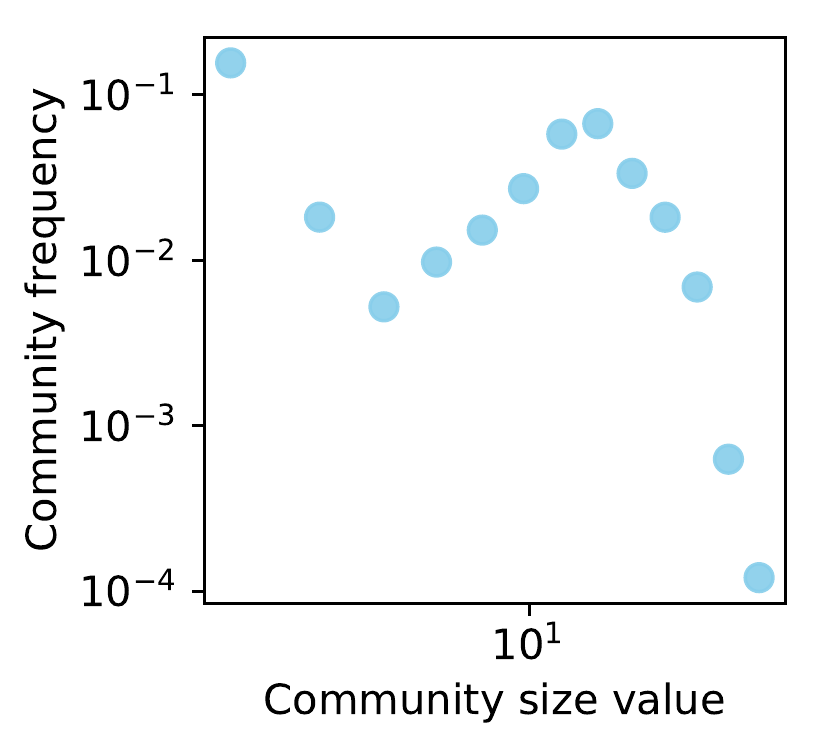} & \includegraphics[width=0.25\textwidth]{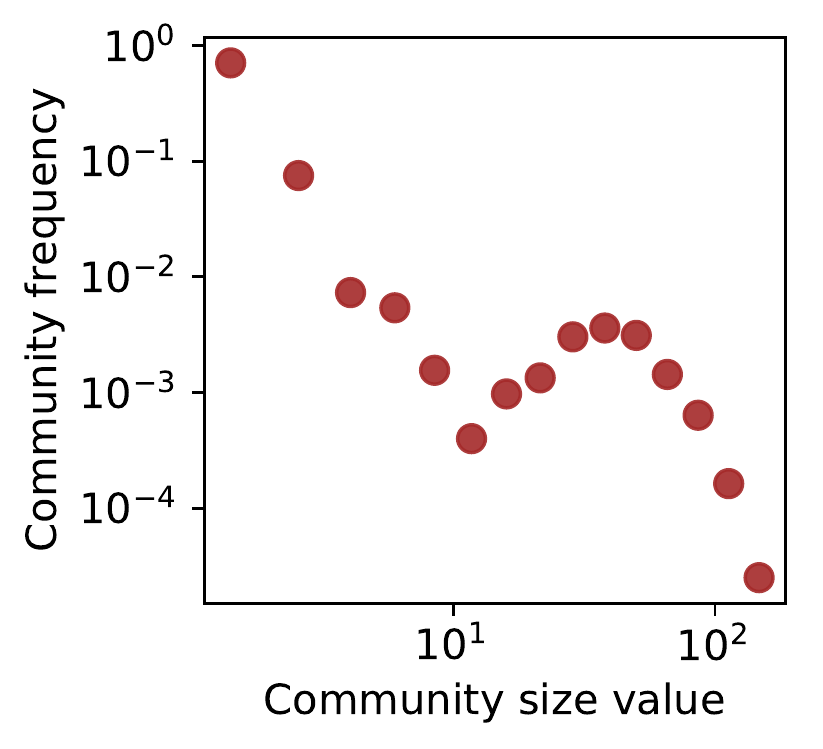} &
\includegraphics[width=0.25\textwidth]{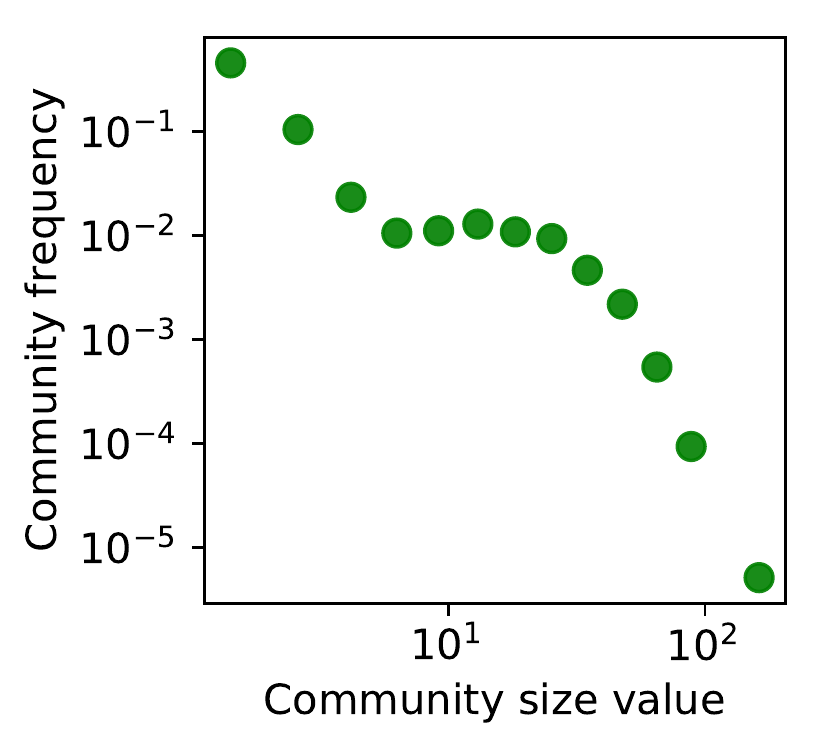}\\
$a$ bin size value & $1.2$ & $2$ or $1.35$ &$2$ or $1.35$ \\
$p$ value & $0.43$ & $0.31$ &$0.81$ or $0.19$ \\
\end{tabular}
\caption{The binned community size distribution for the simulation with nodes $3000$($40000$). After 4 generation (one generation defined as the number of nodes in the networks). The $p$ values are calculated for the Null hypothesis that the to be tested distribution a \textit{power-law} (double log) distributtion. In the case of \textbf{RBRWD} model, the high $p$ value indicates that it is unlikely to reject the loglog distribution hypothesis correctly, that means in the case of \textbf{RRWD} model, we have higher confidence to reject the hypothesis of loglog distribution when the $p$ value is low.}\label{fig:commsize}
\end{figure}

In the work of Lancichinetti~\textit{et al.}, Infomap is the algorithm used to detect communities~\cite{lancichinetti2010characterizing}. We decide to follow their work and apply the Infomap algorithm. In \figref{fig:commsize}, original graphs have no clear community structures; after $2$ to $4$ generations, the scaling of community sizes starts to emerge, and the distributions evolve to no~-Gaussian and power-law like distributions.

%In the networks of shareholder, we can observe that the largest connected component also have the scaling behaviours of degrees and communities sizes.
% One may argue that Infomap algorithms based on random walks, communities will emerge naturally if the model is based on random walks, however, it will not give rise to robust scaling of communities. (\textbf{Qing's comments:} Mayb it is better not to mention it ?)

% ---------------------------------------------
\subsection{Percolation problems}
% Recall the percolation statistics we measured for the shareholder network~\cite{yao2018shareholder}.
We have applied percolation analysis\footnote{You can think it as cluster analysis.} to understand the different node types' topological positions. The measurement starts with a connected component and is defined for different node types. We remove the chosen type of node one by one at random. The number of connected components is measured each time one node is removed. The simulation results are averaged over $100$ realisations. This measure has been used in the shareholder networks~\cite{yao2018shareholder}.
In Turkey and Netherlands' shareholder networks, removing different types of investors results in different rates of change in the number of connected components.

\begin{figure}
    \centering
    \includegraphics[scale=0.9]{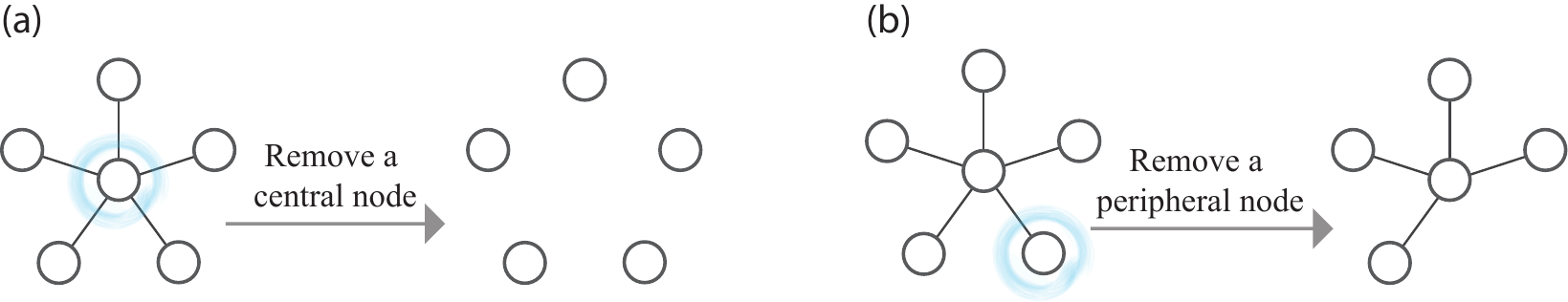}
    \caption{The illustration for percolation analysis, starting with a single component. (a) Removing a node at centre position will result in $5$ components. (b) Removing a node at non-center position will result in $1$ component. Nodes taken are at center will make the network break into many small components quickly while nodes taken from the periphery of the network will typically leave it is a single component.}\label{fig:illu_percolation}
\end{figure}

When removing nodes, the disappearance of the associated edges might break the connected graph into various components. Hence, the number of components may increase. Core nodes at positions linking components together break the graph into many small components quickly while peripheral nodes connected by others are not able to break the graph into many smaller components, see a star graph as an example in~\autoref{fig:illu_percolation}.

\begin{figure}[h!]
    \centering
    \includegraphics[scale=0.9]{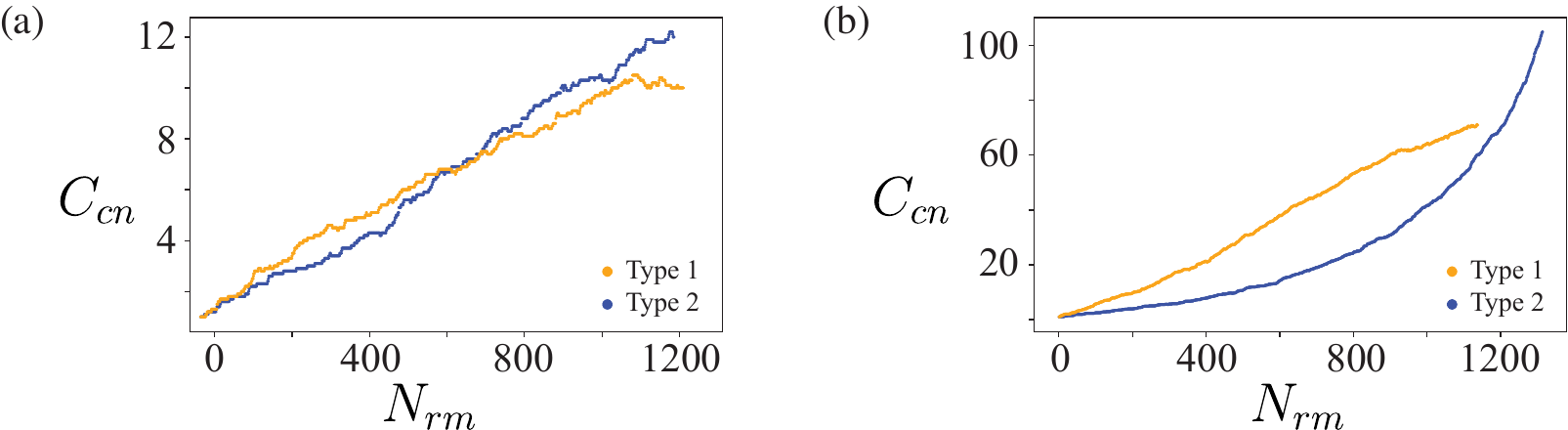}
    \caption[The percolation plots of random walk models]{The percolation plots of random walk models. The $N_{nr}$ on the $x$-axis is the number of nodes removed and $N_{cc}$ on $y$-axis is the number of connected components. The entire directed graph has $3000$ nodes and $14914$ edges. (a) is for random walks regardless of node types (largest connected component has $2451$ nodes and $14635$ edges) and (b) is for random walks considering nodes types (largest connected component has $2470$ nodes and $111255$ edges). Note: the edges are preserved in directed graph, however is not preserved in projected network.}\label{fig:percolation}
\end{figure}

The rate of percolation reflects a node type's topological position. And we apply this measurement on the projected graph $\mathcal{P}$ for both unbiased (\textbf{RRWD}) and biased (\textbf{RBRWD}) models.

In \autoref{fig:percolation} (a) for the \textbf{RRWD} model, if we perform random walks regardless of types (we just label them), the nodes with randomly assigned types will be at similar topological positions. Removing different types of nodes results in similar percolation behaviours. The `Type $1$' and `Type $2$' behave identically. Merely removing one type of nodes does not dismantle the largest connected component. That results in a small increase in the number of components.

However, In ~\autoref{fig:percolation} (b) (\textbf{RBRWD}), in projected networks created by labelled random walks, the nodes of different types will be at different topological positions. Type $1$ has a shallower slope at the beginning when removing the nodes and then followed by a steeper and steeper rate of the increase of components.

The reason for the shallower slope of Type $1$ at the beginning of the percolation analysis is that when starting removing nodes, the Type $1$ nodes are initially tangled with Type $2$ and most of Type $1$ are at core positions while only a few Type $2$ nodes are at core positions, and most of them are at peripheral positions. When removing a small amount of Type $1$, the giant components dismantle slower at the beginning because of some Type $2$ or other Type $1$ linking the nodes together. After removing a considerable amount of Type $1$, the linking nodes are removed, and the components break into small pieces quickly.

The reason of the higher slope of Type $2$ in~\autoref{fig:percolation} (b) at the small $N_nr$ is that some of them are in the centre and glue the components together, similar to the `weak' ties. In the real network, we have fewer Type $1$ nodes only at the core positions linking the components, and the network falls into small pieces quickly similar to the second stage of Type $1$ in ~\autoref{fig:percolation} (b).

Additionally, even though we preserved the number of edges in the directed graph, the number of edges in the projected network for unbiased random walks (regardless of node types) is much larger than for random walks depending on node types. The reason for this is because for a random walker, in the directed graph, there is a higher probability to rewire to nodes with higher degrees. Then in projected networks, the nodes linked to large degree nodes create dense cliques which are the main contribution to a larger number of degrees for the random walk regardless of types. That the more densely the networks are, the harder to be broken into small components explains the difference of scale on the y-axises in~\autoref{fig:percolation} (a) and (b).

% ------------------------------------------------------
\subsection{Average shortest path length}

\begin{figure}[!h]
    \centering
    \includegraphics[scale=0.9]{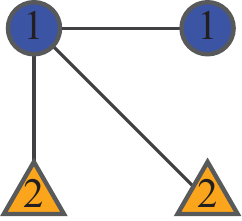}
    \caption[The illustration for average shortest path]{The illustration for average shortest path. One path between two Type $1$ nodes, the length is $1$, $h_{1,1} =1$. One path between two Type $2$ nodes, the length is $2$, $h_{2,2} =2$. $4$ paths between Type $1$ and Type $2$ nodes, the length is $1$ and $2$, $h_{1,2} = \frac{1+1+2+2}{4} = 1.5$.}\label{fig:illu_shortespath}
\end{figure}

We assume that different types of nodes interact differently and have different average shortest paths among different types. Let $\ell(i, j)$ denote the length of the shortest path between node $i$ and node $j$.  We want to look at how the typical length scale varies depending on the node type at the start and end of paths. So we find that the average path length $\overline{{\ell}_{\alpha,\beta}}$ of paths between nodes of type $\alpha$ and $\beta$ where $\alpha,~\beta \in \{$ Type $1$, Type $2$ $\}$ is:
\begin{equation}
    h_{\alpha,\beta} = \sum_{\substack{i,j \\ i\neq j}} \delta(x_i,\alpha) \delta(x_j,\beta)
    \, ,
    \quad
    \overline{\ell_{\alpha,\beta}} :=
    \frac{1}{h_{\alpha,\beta}}
    \sum_{\substack{i,j \\ i\neq j}} {\ell(i,j)}
    \delta(x_i,\alpha) \delta(x_j,\beta)
\end{equation}
where $h_{\alpha,\beta}$ is the number of node pairs $(i, j)$ where $x_i=\alpha$ and $x_j=\beta$. The illustration is in~\autoref{fig:illu_shortespath}

\begin{figure}[!h]
    \centering
    \includegraphics[scale=0.9]{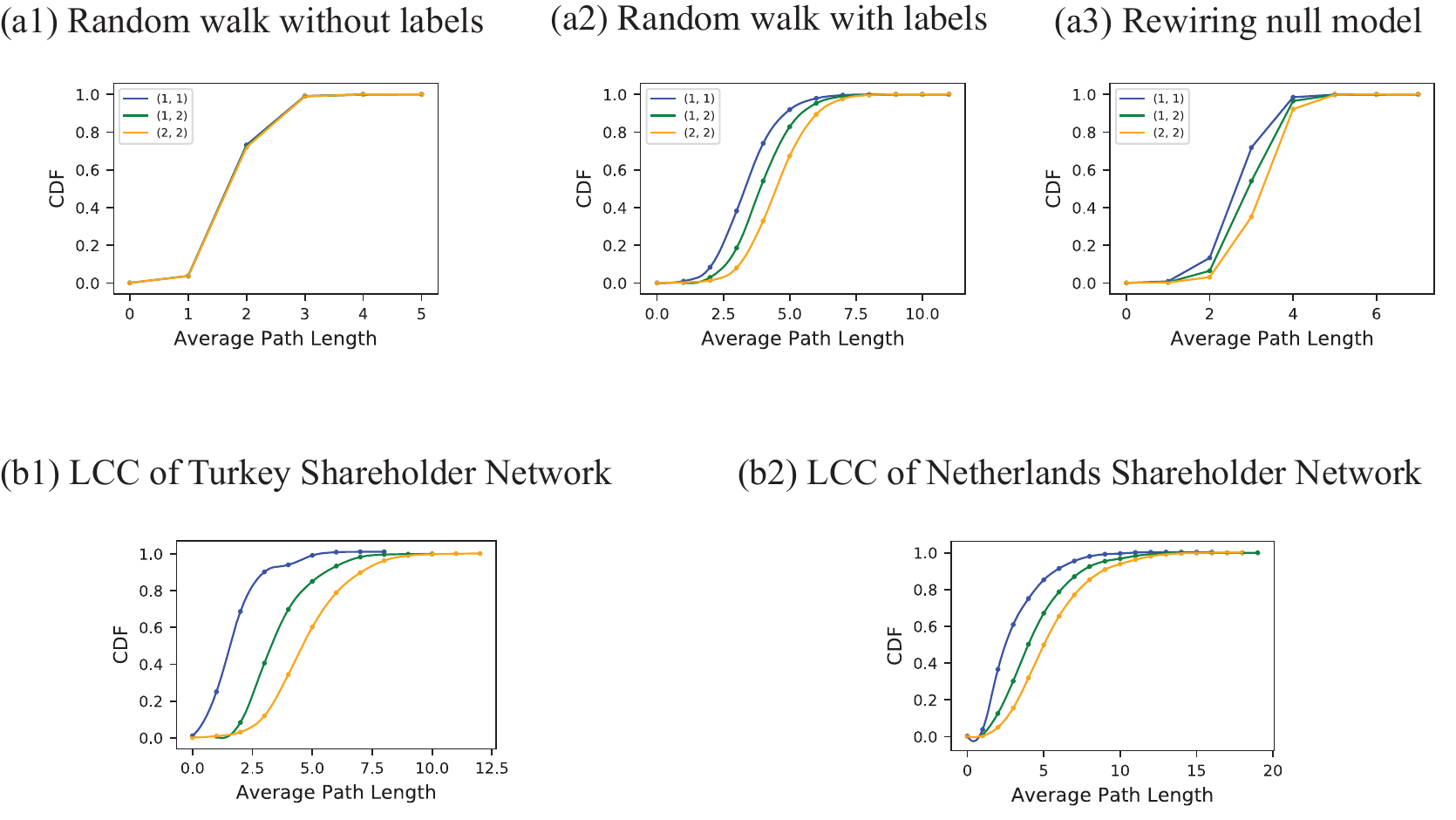}
    %\centering
    % \begin{tabular}{cc}
    % (a.1) Random walk without labels & (a.2) Random walk with labels \\

    % \includegraphics[scale=1]{5_model_fig/cdf_simulation_unlabel.pdf}     &
    % \includegraphics[scale=1]{5_model_fig/cdf_simulation.pdf} \\
    % (a.3)Rewiring null model & The simulation lcc 14046 edges and 2468 nodes \\
    % \includegraphics[scale=1]{5_model_fig/cdf_simulation_null.pdf} & \\
    % (b.1) LCC of Turkey Shareholder Network & (b.2) LCC of Netherlands Shareholder Network  \\
    % \includegraphics[scale=1]{5_model_fig/cdf_tk.pdf} &
    % \includegraphics[scale=1]{5_model_fig/cdf_nl.pdf} \\
    % 1299 nodes and 11616 edges & 3152 nodes and 120935 edges \\
    % \end{tabular}
    \caption[The cumulative density plots for average shortest path lengths between different types of nodes in LCC simulated data]{The cumulative density plots for average shortest path lengths between different types of nodes in LCC simulated data, (a1) is for the networks evolving through random walk without labels and (a2) is for the networks evolving through random walk with labels. (a3) is for the networks obtained after double edge rewiring the networks of (a2). The blue line is for $\bar \ell_{11}$ and orange line is for $\bar \ell_{22}$(which is equal to $\bar \ell_{21}$ in an undirected graph) and green one is for $\bar \ell_{12}$. Type $1$ is with no controlling preference while Type $2$ has preference to attaching to nodes with no more than $2$ in-degree. In (b1) and (b2) is the shortest path lengths between different types of nodes in LCC of Turkey and Netherlands shareholder networks separately. In (b1), Type $1 = $ Bank, and Type $2 = $ Individuals; In (b2), Type $1 =$ Financials and Type $2 = $Corporate.} %\tcomment{Need to show difference between these model plots top and other models e.g. after pair rewiring ignoring labels or single edge rewiring (ER model).}}
    \label{fig:length_distribution}
\end{figure}

\autoref{fig:length_distribution} shows the comparison of the average shortest path length for types between the real data and the models. In general, real data has an average shortest path of $4$, which is longer than the simulated graph. The possible explanation would be that the size of the simulated graph is much smaller than the real graph. Nevertheless, it is satisfactory that the model with labels can mimic the qualitative behaviour of the real-world data of the shareholder networks. In Turkey, Banks appear to be Type $1$, while Families are Type $2$. In the Netherlands, Financials seem to be of Type $1$ while corporates are Type $2$.

%The blue lines in Figure~\ref{tab:length_distribution} in (a.2), (b.1) and (b.2).

% \section{Simulation Motif Transition density}

% \begin{figure}[h!]
%     \centering
%     \includegraphics[width = \textwidth]{transtion2to4}
%     \caption{Motif transition density 2 to 4}
%     \label{f24t}
% \end{figure}

% \begin{figure}[h!]
%     \centering
%     \includegraphics[width = \textwidth]{transtion4to6}
%     \caption{Motif transition density 4 to 6}
%     \label{f46t}
% \end{figure}
% \begin{figure}[h!]
%     \centering
%     \includegraphics[width = \textwidth]{2-4Ev}
%     \caption{Spectral Density of eigenvalue 2-4}
%     \label{f24t}
% \end{figure}

% \begin{figure}[h!]
%     \centering
%     \includegraphics[width = \textwidth]{4-6Ev}
%     \caption{Spectral Density of eigenvalue 4-6}
%     \label{f24t}
% \end{figure}
% \begin{figure}[h!]
%     \centering
%     \includegraphics[width = \textwidth]{2-4cor}
%     \caption{Correlation Matrix of from 2-4}
%     \label{f24t}
% \end{figure}
% \begin{figure}[h!]
%     \centering
%     \includegraphics[width = \textwidth]{4-6cor}
%     \caption{Correlation Matrix of from 4-6}
%     \label{f24t}
% \end{figure}

\section{Discussion}

Motivated by the real-world observation, we build a model biased random walk rewiring process. The main advantage of the random walk model is to mimic the real world's human or organisation's behaviours via local researching.
The phenomenon of the scaling of community size emerges from the local interaction process is qualitatively reproduced by our model. So is the analysis of percolation and the average shortest path length depending on the types.
The simple models do not have these properties. Our model corroborates our anticipation that types are essential when understanding the forming of networks, especially for shareholder networks. We further expect this model capturing heterogeneous behaviours can describe more types of networks.
Another feature of this model which is different from the previous models is that it generated small components and isolated nodes, not just one single connected component. It is a more realistic picture of the real world: not everyone connects to everyone. The randomness in the model provides a mechanism to connect the disconnected pieces.

The projection we used in the model is the representation of sharing risk through the common targets. In recent work, \cite{bourles2017altruism} has applied the theory on networks to model risk sharing in the fixed networks.
In their work, the network of perfect altruism is connected by lots of nodes. That means with a high degree, the individual in the centre of (shorter average path length) network is better insured for the small shocks~\cite{bourles2017altruism}. In our model, the nodes performing random walks without control, are those who are willing to share risk with others and be at this insured position.

\begin{figure}
    \centering
    \includegraphics[scale=0.7]{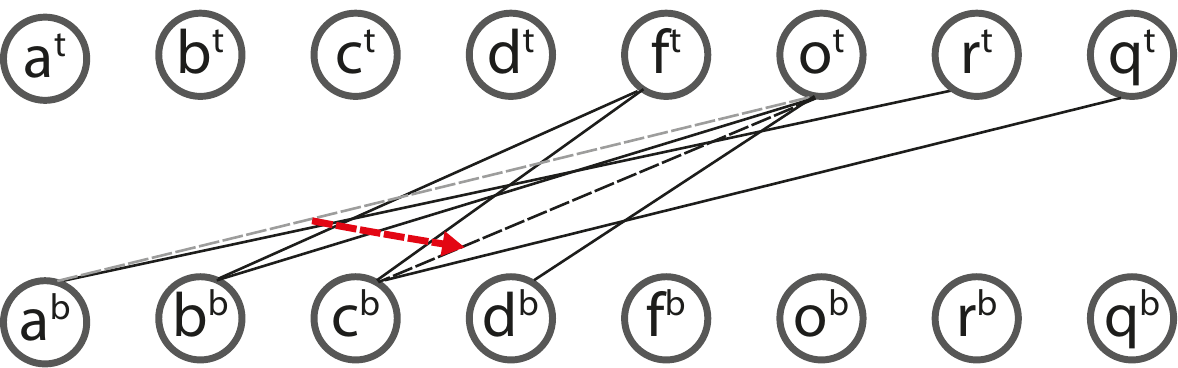}
    \caption[Illustration of corresponding bipartite graph]{Illustration of corresponding bipartite graph of the directed graph where the model performs the random walk rewiring.}
\end{figure}

In our models, once a directed initial random graph is created, we fix the number of nodes and edges of that directed graph. In the real world, new nodes will enter, and some nodes will exit the networks, which have not been captured in this work. If starting with a denser graph, it will not generate scaling of community size distributions. Apart from the number of nodes $|\vertex|$ and edge creation probability $p_{er}$ in the directed graph, another parameter in this model is the number of types and the threshold to separate the preferences of different types. In a future study, we can introduce more two types of nodes (the types of random walks), modify the threshold $\theta_{e}$ of two types, or use a smoother function to separate the preferences.
Again, this research qualitatively matches the statistics of shareholder networks -- more work to be done to extract the parameters to reproduce the real-world systems systematically.

\clearpage

\bibliographystyle{unsrt}
%\bibliographystyle{abbrvnat}
%\bibliographystyle{naturemag}

% For natbib?
%\bibliographystyle{agsm}
%\bibliographystyle{apsrr}
%\bibliographystyle{dcu}

% Local File
\bibliography{main}

\end{document}